\documentstyle[12pt]{article}

\global\arraycolsep=1pt
\begin{document}

\renewcommand{\thefootnote}{\fnsymbol{footnote}}
\begin{titlepage}
\begin{flushright}
         NYU.Th.Ph.97-22\\
October 10, 1997
\end{flushright}
\begin{center}
{\Large \bf Continuum and Lattice \\ Coulomb-Gauge Hamiltonian}
\footnote{Lectures given at the NATO ASI on Confinement, Duality and
Non-perturbative Aspects of QCD, Cambridge, UK, June 23 - July 4, 1997}
\lineskip .75em
\vskip 3em
\normalsize

\vskip 1 em
{\large Daniel Zwanziger}\footnote{email address:
daniel.zwanziger@nyu.edu}\footnote{Research supported in part by the
National Science Foundation under grant no. PHY-9520978}
\\ {\it  Physics Department, New York University, New York, NY 10003, USA}
\end{center}
\vskip 1 em
\begin{abstract} 
	We review the canonical quantization of
continuum Yang-Mills theory, and derive the continuum Coulomb-gauge
Hamiltonian by a simplification of the Christ-Lee method.  We then
analogously derive, by a simple and elementary method, the lattice
Coulomb-gauge Hamiltonian in the minimal Coulomb gauge (and in other
Coulomb gauges) from the known Kogut-Susskind Hamiltonian.
\end{abstract}

\end{titlepage}
\renewcommand{\thefootnote}{\arabic{footnote}}
\setcounter{footnote}{0}


 
\newcommand{\ip}[2]{(#1, #2)}


\section{Introduction}      

        In the Hamiltonian formalism one can calculate the spectrum
directly \cite{Luescher, Cutkosky, vanBaal, vdHeuvel}.  In the Coulomb
gauge, the color form of Gauss's law, which would appear to be essential
for confinement, is satisfied exactly.  Moreover an obvious confinement
mechanism suggests itself in the Coulomb gauge namely, a long-range
instantaneous color-Coulomb potential.

		This raises the question, can the Coulomb gauge be regularized? 
Certainly, on the lattice any configuration may be fixed to the Coulomb
gauge.  The lattice Coulomb hamiltonian which will be described in this
lecture is an implementation of the Coulomb-gauge dynamics with lattice
regularization.  As to whether the Coulomb gauge is perturbatively
renormalizable in the continuum theory, the question is under active
investigation \cite{Z1, BZ}.  The answer appears to be yes. 

\section{Classical Yang-Mills Equations} 

	Classical Yang-Mills theory is designed to be invariant under the group
of local gauge transformations $g(x)$.  We shall consider the local
SU(N) theory, $g(x) \in SU(N)$.  The quark field transforms covariantly,
$\psi(x) \rightarrow ^g \! \! \psi(x) = g(x) \psi(x)$.  The
gauge-covariant derivative $D = D(A)$, defined by
$(D_\mu \psi)(x) = [\partial_\mu + t^a A_\mu^a(x)] \psi(x)$ 
also transforms covariantly, 
$(D_\mu \psi)(x) \rightarrow g(x) (D_\mu \psi)(x)$,
provided that the connection $A$ transforms according to
\begin{equation}
A_\mu(x) \rightarrow ^g \! \! A_\mu(x) 
= g(x) \partial_\mu g^\dagger(x) + g(x) A_\mu(x) g^\dagger(x) . 
\label{eq:lgt}
\end{equation}
Here the anti-hermitian matrices $t^a$ form
a representation of the Lie algebra of SU(N), 
$[t^a, t^b] = f^{abc} t^c$, and $A_\mu \equiv t^a A_\mu^a$.  The {\em
local gauge principle} states that $A$ and $^g \! A$ are physically
identical $A \sim ^g \! \! A$.  The physical configuration space 
$\cal Q = A/G$
is the space of connections ${\cal A} = \{ A \}$ modulo the local gauge
transformations ${\cal G} = \{ g \}$.  This space is
non-trivial because the local gauge transformation is
non-linear.

	The Yang-Mills field tensor,
\begin{equation}
F_{\mu \nu} = \partial_\mu A_\nu - \partial_\nu A_\mu
                + [A_\mu, A_\nu] = t^a F_{\mu \nu},
\end{equation}
 transforms  homogeneously, 
\begin{equation}
F_{\mu \nu}(x) \rightarrow ^g \! \! F_{\mu \nu}^a(x) 
= g(x) F_{\mu \nu}(x) g^\dagger(x) . 
\end{equation}
The Yang-Mills action
\begin{equation}
S_{\rm YM} = g_0^{-2} \int d^4x (1/4) ( - g^{\mu \kappa} g^{\nu \lambda} )
        F_{\mu \nu}^a F_{\kappa \lambda}^a,
\end{equation}
is gauge-invariant and yields the field
equations 
$D^\mu F_{\mu \nu} \equiv \partial^\mu F_{\mu \nu} + [A^\mu, F_{\mu \nu}]
= 0$.  Here the metric tensor is defined by 
$g_{\mu, \nu} = {\rm diag}(-1, 1, 1, 1)$,  and $g_0$ is the coupling
constant.  (If quarks were present we would have 
$D^\mu F_{\mu \nu} = - j_{{\rm qu,}\nu}$.)  
	
	The color-electric and magnetic fields are defined by
$E_i \equiv F_{i 0} = - \dot{A}_i + D_iA_0$ and 
$B_i = F_{jk}$ for $i, j, k$ cyclic.  

\section{Gauss's law and Color-Coulomb Potenial} 

The time derivative of
$A_0$ nowhere appears in the preceding equations, and the field equation
which corresponds to variation with respect to $A_0$ is the
time-independent constraint which constitutes the
color Gauss's law,  
\begin{equation}
D_i E_i \equiv \nabla_i E_i + A_i \times E_i = 0 \; \; \; {\rm (or} =
\rho_{\rm qu} {\rm )},
\end{equation} 
where $(A \times E)^a \equiv f^{abc} A^b E^c$.  This law
may be used to fix the longitudinal part of the color-electric field.  

	We write
\begin{equation}
E_i = E_i^{\rm tr} - \nabla_i \Omega.
\end{equation}
where $\nabla_i E_i^{\rm tr} = 0$, and $\Omega$ is the {\em
color-Coulomb potential}.  We have
\begin{equation}
D_i E_i   
= - D_i \nabla_i \Omega + A_i \times E_i^{\rm tr} = \rho_{\rm qu}.
\end{equation} 
The three-dimensional Faddeev-Popov operator is defined by
\begin{equation}
	M = M(A) \equiv - D_i(A) \nabla_i
\end{equation}
Gauss's law fixes the color-Coulomb potential 
\begin{equation}
 \Omega(x) = (M^{-1} \rho)(x) = \int d^3y \; M^{-1}(x, y; A) \rho(y).
\end{equation} This equation, which appears to be neither Lorentz nor
gauge non-invariant, in fact holds in every Lorentz frame and every
gauge.  Here
$\rho$ is a color-charge density
\begin{equation}
	\rho \equiv - A \times E^{\rm tr} + \rho_{\rm qu}
\end{equation}
which does not contain the complete color-charge density of the gluons,
$\rho_{\rm gl} = - A \times E$, but only the part that comes from
$E^{\rm tr}$.  

	The color-Coulomb potential is propagated by the Green's function
\begin{equation}
	M^{-1} = (- D_i \nabla_i)^{-1} 
	= ( - \vec{\nabla}^2 - A_i \times \nabla_i)^{-1}
\end{equation} 
of the three-dimensional Faddeev-Popov operator.  The famous
anti-screening property of non-Abelian gauge theory arises from this
circumstance.  For whereas $(- \vec{\nabla}^2)$ is a positive operator,
the second term $(- A_i \times \nabla_i)$ may be positive or negative. 
In the latter case there is a cancellation in the denominator which
produces an enhanced color-Coulomb potential.

\section{Classical canonical formalism} 

	The Yang-Mills lagrangian is of the form
\begin{equation}
L_{\rm YM} = \int d^3x {\cal L}_{\rm YM}
\end{equation}
where the Lagrangian density is 
\begin{equation}
{\cal L}_{\rm YM} = g_0^{-2} (1/2) ( \vec{E}^2 - \vec{B}^2 ) .
\end{equation}
The time derivatives are contained in $E_i = - \dot{A}_i + D_iA_0$, and
the canonical momentum densities are given by
\begin{eqnarray}
\pi_i & = & \partial {\cal L} / \partial \dot{A}_i = - g_0^{-2} E_i 
\nonumber  \\
\pi_0 & = & \partial {\cal L} / \partial \dot{A}_0 = 0 .
\end{eqnarray}
The classical Hamiltonian
\begin{eqnarray}
H & = & \int d^3x ( \pi_i \dot{A}_i - {\cal L})  \nonumber \\
H & = & \int d^3x \; [ \; (1/2) ( g_0^2 \: \vec{\pi}^2 
      + g_0^{-2} \vec{B}^2 ) - A_0 D_i \pi_i \; ]
\end{eqnarray}
produces the equations of motion
\begin{eqnarray}
\dot{A}_i = \delta H / \delta \pi_i  & \Rightarrow & \pi_i = - g^{-2} E_i
\nonumber  \\
\dot{\pi}_i = - \delta H / \delta A_i  & \Rightarrow & 
D_0E_i + \epsilon_{ijk}D_jB_k = 0,
\end{eqnarray}
and the Gauss's law constraint
\begin{equation}
0 = \delta H / \delta A_0 \Rightarrow D_i \pi_i = 0
\end{equation}

\section{Quantization in the Weyl gauge}

	Before deriving the lattice Coulomb Hamiltonian, we shall derive the
Christ-Lee Coulomb Hamiltonian for continuum gauge theory which it
resembles.  Christ and Lee start by canonical quantization in the Weyl
gauge, $A_0 = 0$.  This is not a complete gauge-fixing and allows
time-independent local gauge transforamtions $g = g(\vec{x})$.  In this
gauge we have $E_i = -
\dot{A}_i$, and the previous Hamiltonian reduces to
\begin{equation}
H_W = \int d^3x \; (1/2) ( \; g_0^2 \: \vec{\pi}^2 
      + g_0^{-2} \vec{B}^2  \; ) .
\end{equation}
This Hamiltonian generates the previous equations of motion in the Weyl
gauge, but not the Gauss's law constraint $D_i\pi_i = 0$.  In the
classical theory that is imposed as an initial condition that holds at 
$t = 0$, and which is preserved by the equations of motion.

	In the Christ-Lee method, one quantizes in the Weyl gauge by treating $A$
and $\pi$ as Cartesian variables, with canoncal equal-time commutation
relations
\begin{equation}
[A_i^a(\vec{x}), A_j^b(\vec{y})] =  
[\pi_i^a(\vec{x}), \pi_j^b(\vec{y})] = 0    
\end{equation}
\begin{equation}
 [\pi_i^a(\vec{x}), A_j^b(\vec{y})] 
 =  - i \delta^{ab} \; \delta_{ij} \; \delta(\vec{x} - \vec{y}) .
\end{equation}
There is then no ordering problem in the Weyl Hamiltonian, nor in the
definition of the left-hand side of Gauss's law
\begin{equation}
G^a(x) \equiv (D_i \pi_i)^a(x) 
= \nabla_i \pi_i^a + f^{abc}A_i^b(x) \pi_i^c(x) .
\end{equation}

	Moreover with the above canonical commuation relations one observes that
$G^a(x)$ is the generator of local time-independent gauge  
transformations,
\begin {equation}
[G^a(x), \pi_i^b(y)] = i \delta(x - y) f^{abc} \pi_i^c(y) \label{eq:gtpi}
\end{equation}
\begin {equation}
[ G^a(x), A_i^b(y) ] = - i D_i^{ab} \delta(x - y),
\end{equation} 
under which $\pi$ transforms homogeneously.  The $G^a(x)$ satisfy the
Lie algebra of the time-indpendent local gauge group, 
\begin{equation}
[G^a(x), G^b(y)] = i \delta(x - y) f^{abc} G^c(y)
\end{equation}
and moreover these local time-independent gauge transformations are a
symmetry of the Hamiltonian $[ G^a(x), H_W ] = 0$.  Therefore it is
consistent to impose Gauss's law as a subsidiary condition which must be
satisfied for a wave function $\Psi = \Psi(A)$ to represent a physical
state,
\begin{equation}
G^a(x) \Psi = 0.    \label{eq:subsidiary}
\end{equation}
This condition is also the statement that the wave function is
invariant under time-independent local gauge transformations.

	The inner product is defined by the functional integral
\begin{equation}
(\Phi, \Psi) = \int dA \; \Phi^{*}(A) \Psi(A),
\end{equation}
and the Hamiltonian by $H_W = T + V$, where $T$ and $V$ correspond
respectively to kinetic
(electrical) and potential (magnetic) energies.  Here 
\begin{equation}
V = \int d^3x \; (1/2) g_0^{-2} [B_i^a(x)]^2 
\end{equation}
acts by multiplication, and we define $T$ by its expression as a
quadratic form
\begin{equation}
(\Phi, T \Psi) = \int dA \int d^3x \; (g_0^2/2) 
  \; \delta \Phi^*(A) / \delta A_i^a(x) 
\; \delta \Psi(A) / \delta A_i^a(x) .
\end{equation}

\section{Minimal continuum Coulomb gauge}

	In the Christ-Lee method \cite{CL}, one solves the Gauss's law subsidiary
condition explicitly to obtain the Hamiltonian in the physical subspace.
In the original Christ-Lee paper, this is done by choosing coordinates in
$A$-space adapted to the gauge orbit.  A gauge orbit is a set of
gauge-equivalent configurations $A \sim \: ^g \! A$.  (Here we refer only
to time-independent local gauge transformations $g$.)  This may be done by
writing $A = \: ^g \! B$, where $B$ is transverse, $\nabla_i B_i = 0$,
and $g$ runs over the set of all possible gauge transformations.  Here
the transverse configuration $B$ labels the gauge orbit and $g$ labels
the point on the orbit.  One way to address the problem of Gribov copies
\cite{Gribov}, is to restrict $B$ to a
subset $\Lambda$ of transverse configurations $\Gamma$ known as the
fundamental modular region.  For an alternative approach, see
\cite{Friedbergetal}.

	A convenient may way to choose $\Lambda$ is by means of a minimizing
function
\begin{equation}
F_A(g) \equiv \: \parallel \, ^g \! A \parallel^2 
= \int d^3x | g \vec{\nabla} g^\dagger + g \vec{A} g^\dagger |^2.
\end{equation}
The fundamental modular region $\Lambda$ is defined to be the set of $A$
that are an absolute minima of this function,
\begin{equation}
\Lambda \equiv \: \{A: \: \parallel A \parallel^2 \leq 
\: \parallel \, ^g \! A \parallel^2 {\rm for all} \; g \} .
\end{equation}
With suitable assumptions on the space of configurations, one may show
that such a minimum exists \cite{dellAZ}.  This definition (almost)
uniquely fixes a single point on each gauge orbit, to within global gauge
transformations, because the absolute minimum of the functional is unique
apart from ``accidental'' degeneracies.  These are degenerate (equal)
absolute minima of the minimizing functional.  They constitute the
boundary of
$\Lambda$.  The fundamental modular region is obtained from
$\Lambda$ by topologically identifying the boundary points that are
gauge-equivalent degenerate absolute minima.  (For a more extensive
discussion of $\Lambda$ see \cite{Z2}.)

	We may easily derive some properties of the fundamental modular
region.  At a relative or absolute minimum the functional is
stationary to first order and its second-order variation is positive. 
With 
$g = \exp(\omega) = 1 + \omega + (1/2)\omega^2 + ...$ 
an elementary calculation gives
\begin{equation}
\parallel \, ^g A \parallel^2 = \parallel \, A \parallel^2
- 2 ( \nabla_i A_i,  \omega) + (1/2) (\omega, M(A) \omega) + ... \; \; ,
\end{equation} 
where $M(A) = - D_i(A) \nabla_i$ is the three-dimensional Faddeev-Popov
operator.  Since this must be positive for all $\omega$, we obtain the
Coulomb gauge condition
$\nabla_i A_i = 0$. {\em In addition} we find that $M(A)$ is a positive
operator,
$(\omega, M(A) \omega) \geq 0 \; {\rm for all} \; \omega$ and for all
$A \in \Lambda$.  These two conditions define the Gribov region
$\Omega$ which is the set of all relative and absolute minima, of
which the fundamental modular region $\Lambda$ is a proper subset,
$\Lambda \subset \Omega$.  Note that because Faddeev-Popov operator $M(A)$
is a positive operator for $A \in \Lambda$, M(A) is invertible in the
minimal Coulomb gauge and has non-negative eigenvalues.

\section{Continuum Coulomb-gauge Hamiltonian}

	In the present lecture we shall not explicitly change variables from
$A$ to the adapted coordinates $B$ and $g$, as was originally done by
Christ and Lee \cite{CL}, but instead use the more efficient
Faddeev-Popov ``trick'' to obtain $H_{\rm coul}$. 

	The Faddeev-Popov identity reads
\begin{equation}
1 = \int dg \; \delta_\Lambda(\nabla_i \,^g \! A) 
\det[ M( \,^g \!A) ],
\end{equation}
 where $M(A) =  - \nabla_i D_i(A)$ is the Faddeev-Popov operator.  Here
$\delta_\Lambda(\nabla_i A_i)$ is the restriction of 
$\delta(\nabla_i A_i)$ to the fundamental modular region $\Lambda$,
namely
$\delta_\Lambda(\nabla_i A_i) = \delta(\nabla_i A_i) \chi_\Lambda(A)$,
where $\chi_\Lambda(A) = 1$ for (transverse) $A \in \Lambda$, and 
$\chi_\Lambda(A) = 0$ otherwise.

	We require wave functions to be gauge invariant, 
$\Psi(\,^g \! A) = \Psi(A)$, and we wish to express the inner-product  of
such wave-functions
$(\Phi, \Psi)$ as an integral over the parameters
$A^{\rm tr} \in \Lambda$ that label the orbit space.  For this purpose we
insert the Faddeev-Popov identity into the inner product and obtain 
\begin{eqnarray}
(\Phi, \Psi) & = & \int dA \; \Phi^{*}(A) \Psi(A)  \nonumber \\
& = & \int dA dg \; \delta_\Lambda(\nabla_i \,^g \! A_i) 
\det[ M( \,^g \! A) ]  \; \Phi^{*}(A) \Psi(A)   \nonumber  \\
& = & \int dA dg \; \delta_\Lambda(\nabla_i \,^g \! A_i) 
\det[ M( \,^g \!A) ]  \; \Phi^{*}( \,^g \!A) \Psi( \,^g \!A)
\nonumber  \\
& = & \int dA' dg \; \delta_\Lambda(\nabla_i A'_i) 
\det[ M(A') ]  \; \Phi^{*}(A') \Psi(A') \nonumber  \\
& = & N \int dA \; \delta_\Lambda(\nabla_i A_i) 
\det[ M(A_i) ]  \; \Phi^{*}(A) \Psi(A) \nonumber  \\
& = & N \int_\Lambda dA^{\rm tr} \; 
\det[ M(A^{\rm tr}) ]  \; \Phi^{*}(A^{\rm tr}) \Psi(A^{\rm tr}).
\end{eqnarray}
Here $N = \int dg$ is the (infinite) volume of the gauge orbit which is
an irrelevant normalization constant, and $\int_\Lambda dA^{\rm tr}$ is
the functional intregral over all transverse configurations in $\Lambda$.
We have used the invariance of the measure $dA$ under local gauge
transformation, $dA = dA'$ for $A' = \,^g \!A_i$.  These manipulations
are somewhat formal in the continuum theory, but are well defined in
lattice gauge theory.

	We next express the Hamiltonian $H_{\rm W} = T + V$ on the reduced
space of wave functions that depend on $A^{\rm tr}$.  Because 
$V(A)$ is gauge invariant, it acts on the reduced space as the
operator of multiplication by $V(A^{\rm tr})$.

	The expression for $T$ as a quadratic form is particularly convenient
for our purpose because it only involves first derivatives, and this
allows us to apply Gauss's law.  We have
seen (\ref{eq:gtpi}) that 
$\pi_i^a(x) = -i \delta / \delta A_i^a(x)$ transforms homogeneously under
gauge transformations, so for gauge-invariant wave functions
the quantity
$\delta \Phi^* / \delta A_i^a(x) \; \delta \Psi / \delta A_i^a(x)$ is also
gauge-invariant.  Consequently when we insert the Faddeev-Popov
identity into the quadratic form that defines $T$, we obtain, as for the
inner product,
\begin{eqnarray}
(\Phi, T \Psi) & = & N \int dA \;  \delta_\Lambda (\nabla_i A_i) 
\det[ M(A) ]  \;      \nonumber  \\
 & \times & 
\int d^3x \; (g_0^2/2) 
  \; \delta \Phi^*(A) / \delta A_i^a(x) 
\; \delta \Psi(A) / \delta A_i^a(x) .
\end{eqnarray}
	
	To proceed, we shall use Gauss's law 
$ D_i \delta \Psi /\delta A_i = 0$ 
to express
$\delta \Psi /\delta A_i$ as a derivative with respect to $A^{\rm tr}$. 
For this purpose we write 
$E_i = E_i^{\rm tr} - \nabla_i \Omega$, where 
$E_i = i \delta / \delta A_i$ and 
$\nabla_i E_i^{\rm tr} = 0$.  Here $\Omega$ is the color-Coulomb potential
operator, and
$E^{a, {\rm tr}} = i \delta / \delta A_i^{a, {\rm tr}}$.
With $D_iE_i = \nabla_iE_i + A_i \times E_i$, Gauss's law, reads
\begin{equation}
 - D_i \nabla_i \Omega \Psi = \rho \Psi .
\end{equation}
where we have introduced the color-charge density operator corresponding
to the dynamical degrees of freedom
\begin{equation}
 \rho \equiv - A_i \times E_i^{\rm tr} ,
\end{equation}
As in the classical theory, we may solve for $\Omega \Psi$ by
inverting the Faddeev-Popov operator
$M(A) =  - D_i(A) \nabla_i$.  We have
$\Omega \Psi = M^{-1} \rho \Psi$, 
which gives
\begin{equation}
E_i \Psi= E_i^{\rm tr} \Psi 
- \nabla_i M^{-1}(A) \:( - A_i \times E_i^{\rm tr} ) \Psi .
\end{equation}
Only derivatives with respect to $A^{\rm tr}$  now appear in the
expression for the quadratic form $T$,
\begin{eqnarray}
(\Phi, T \Psi) & = & N \int_\Lambda dA^{\rm tr} \;    
\det[ M(A^{\rm tr}) ]  \;      \nonumber  \\
 & \times & 
\int d^3x \; (g_0^2/2) 
  \; (E_i^a(x)\Phi)^*(A^{\rm tr})
\; (E_i^a(x)\Psi)(A^{\rm tr})  ,
\end{eqnarray}
where $E\Psi$ and $E\Phi$ are defined by the preceding equation. 
The transverse and longitudinal parts of the electric field contribute
separately to the kinetic energy,
\begin{eqnarray}
(\Phi, T \Phi) & = & N \int_\Lambda dA^{\rm tr} \;    
\det[ M(A^{\rm tr}) ]  \;  \int d^3x \; (g_0^2/2)     \nonumber  \\
 & \times & 
[ \; |E_i^{{\rm tr}, a}(x)\Phi |^2
+  |(\nabla_i M^{-1} \rho)^a(x)\Phi |^2 \; ].
\end{eqnarray}

	We have expressed $T$ as a quadratic form on functions
of transverse configurations, $A^{\rm tr} \in \Lambda$ that parametrize
the orbit space, and $V$ as the operator of multiplication by 
$V(A^{\rm tr})$.  These expressions define the Hamiltonian in the
minimal Coulomb gauge, $H_{\rm coul} = T + V$, defined by the gauge
choice $A_i = A_i^{tr} \in \Lambda$.  

\section{Kogut-Susskind lattice Hamiltonian}

In a Euclidean lattice theory the Hamilton $H$ may be derived from the
partition function $Z$.  The partition function is expressed in terms of
the transfer matrix $T$ by  
$Z = \lim_{N \rightarrow \infty} {\rm tr} T^N$, where $N$ is the number of
Euclidean ``time'' slices.  One chooses an asymmetric lattice with
lattice unit $a_0$ in the Euclidean time direction and lattice unit $a$
in spatial directions.  Then 
$T = \lim_{a_0 \rightarrow 0} \exp( - a_0 H)$.  

	In \cite{Z2} the lattice
Coulomb-gauge Hamiltonian was derived directly from this formula, where
the configurations were fixed to the lattice Coulomb gauge.  In the
present lecture we present a much briefer and elementary derivation.   We
shall start with the known Kogut-Susskind Hamiltonian \cite{KS}, $H_{\rm
KS}$, which is the lattice analog of the Weyl Hamiltonian.  We
then fix the gauge, and solve the lattice form of the Gauss's law
constraint to obtain the lattice Coulomb-gauge Hamiltonian on the
reduced space, just as we did in the continuum theory.  

	We designate points on the three-dimensional periodic cubic
lattice by three-vectors with integer components, which we designate
$x, y, ...$ .  Gauge transformations are site variables, 
$g_x \in SU(N)$.  The basic variables of the theory are the link
variables $U_{(xy)} \in SU(N)$ defined for all links (xy) of the lattice,
with $U_{(yx)} = U_{(xy)}^\dagger$.  Local gauge transformations are
defined by 
$U_{(xy)} \rightarrow \, ^gU_{(xy)} = g_x U_{(xy)}g_y^\dagger$.

	The $U_0 = 1$ gauge is the lattice analog of the Weyl $A_0 = 0$ gauge. 
In this gauge, the Hamiltonian obtained from Wilson's lattice action and
the formula $T = \lim_{a_0 \rightarrow 0} \exp( - a_0 H)$ is
\begin{equation}
H_{\rm KS} = T + V
\end{equation}
where
\begin{eqnarray}
T & = & \sum_{(xy)} \; (g_0^2/2a) \; J_{(xy)}^2  \nonumber  \\
V & = & (2N/ag_0^2) \sum_p ( 1 - N^{-1} {\rm Re \: tr} U_p)  .
\end{eqnarray}
Here $a$ is the (spatial) lattice unit.  Henceforth we set $a = 1$.  In
the last expression the sum extends over all plaquettes
$p$ of the (spatial) lattice, and $U_p$ is the product of the link
variables
$U_{(xy)}$ on the links around the plaquette $p$.  The potential energy
operator is gauge invariant $V( ^gU) =
V(U)$, as are physical wave functions, $\Psi( ^gU) =
\Psi(U)$. 

	The variables $J_{(xy)}$ are electric field operators associated to
each link of the lattice, defined as follows.  Let the link variable
$U_{(xy)}$ be parametrized by a set of variables $\theta_{(xy)}^\alpha$,
then, for each link $(xy)$,
\begin{equation}
J_{(xy), a} \equiv i \; J_a^\alpha(\theta_{(xy)}) \; 
\partial / \partial \theta_{(xy)}^\alpha
\end{equation}
is the Lie generator of the SU(N) group, that satisfies the commutation
relations
\begin{equation}
[ J_{(xy), a}, J_{(uv), b} ] 
= i \delta_{(xy),(uv)} f^{abc} J_{(xy), c}.
\end{equation}
For the SU(2) group, $T$ would be the Hamiltonian for a set of
non-interacting tops, located on every link of the lattice.  The
potential energy operator $V$ provides an interaction between tops on
the same plaquette.

	Both $T$ and $V$ are gauge invariant, as is the inner product 
\begin{equation}
(\Psi, \Phi) \equiv \int \prod_{(xy)} dU_{(xy)} \Psi^{*}(U) \Phi(U).
\end{equation}
Here $dU_{(xy)}$ is the Haar measure on each link,
\begin{equation}
dU_{(xy)} \equiv \det\psi_{(xy)} \prod_\alpha d\theta_{(xy)}^\alpha ,
\end{equation}
where $\psi(\theta_{(xy)})$ is the inverse of the matrix 
$J(\theta_{(xy)})$,
\begin{equation}
J_a^\alpha(\theta_{(xy)}) \psi_\alpha^b(\theta_{(xy)}) = \delta_a^b .
\end{equation} 

\section{Minimal lattice Coulomb gauge}

We wish to express the hamiltonian as on operator on the reduced
space of gauge orbits.  For this purpose we require a set of parameters
that label the gauge orbits.  A convenient way to identify the gauge
orbits is to introduce a minimizing function, as in the continuum
theory,  
\begin{equation}
F_U(g) \equiv \sum_{x,i} \; 
\{ \: 1 - N^{-1} {\rm Re \; tr} [( \, ^gU)_{x,i} ] \; \}.
\end{equation}
Here we have introduced the notation
$U_{x,i} \equiv U_{(xy)}$ for $y = x + e_i$ where $e_i$ is a unit vector
in the positive $i$-direction.  The gauge-transformed link
variable is given by 
$ (\, ^gU)_{x,i} = g_x U_{x,i} g_{x +\hat{i}}^\dagger$, where we have
written $\hat{i} \equiv e_i$. The fundamental modular region $\Lambda$ is
defined to be the set of configurations $U$ that are absolute minima of
this function,
\begin{equation}
\Lambda \equiv \{U: F_U(1) \leq F_U(g) \; {\rm for \: all} \: g\} .
\end{equation}
With this gauge choice, the link variables $U_{x,i}$ are made as close to
unity as possible, equitably, over the whole lattice.

	The advantage of this method is that we are assured that a minimizing
configuration exists on each gauge orbit, because the minimizing function
is defined on a compact space (a finite product of SU(N) group
manifolds).  An alternative procedure that is sometimes followed is to
posit a gauge condition.  For example with the exponential mapping
$U_{x,i} =
\exp(A_{x,i})$, one may posit the condition 
$\sum_i(A_{x,i} - A_{x-\hat{i}}) = 0$ which is a lattice analog of the
continuum Coulomb-gauge transversality condition.  An obstacle to
this procedure is that one does not know whether every gauge orbit
intersects this gauge-fixing surface.  On the other hand for the purpose
of formal expansions the alternative procedure may be used. 
	
	As in the continuum case, we observe that, at a relative or absolute
minimum, the minimizing function is stationary to first order, and its
second order variation is positive.  In order to conveniently exploit
these properties, we introduce a one-parameter subgroup of the local gauge
group $g_x(s) = \exp(s\omega_x)$, where 
$\omega_x = t^a \omega_x^a$ is anti-hermitian, 
$\omega_x^\dagger = - \omega_x$.  We write 
$F_U(s) \equiv F_U(g(s))$ and we have
\begin{eqnarray}
dF_U(s) / ds = - N^{-1} \sum_{x,i} \;  {\rm Re \; tr}   \nonumber
[\omega_x ( \, ^gU)_{x,i} - ( \, ^gU)_{x,i} \omega_{x+\hat{i}} \: ]  \\ 
= N^{-1} \sum_{x,i} \;  {\rm Re \; tr} \{
[ \omega_{x+\hat{i}} - \omega_x ]( \, ^gU)_{x,i}  \: \} \label{eq:dFds} \\
= - (2N)^{-1} \sum_{x,i} \;  {\rm tr} \{
 \omega_x [ ( \, ^gU)_{x,i} -  ( \, ^gU)_{x,i}^\dagger
-  ( \, ^gU)_{x-\hat{i},i} + ( \, ^gU)_{x-\hat{i},i}^\dagger ] \: \}.
\end{eqnarray}
At a relative or absolute minimum $U$ we have $dF_U(s) / ds|_{s=0} = 0$. 
To express this condition compactly, we introduce the link variables
\begin{equation}
t^a A_{x,i}^a \equiv A_{x,i} \equiv 
(1/2) (U_{x,i} - U_{x,i}^\dagger )_{\rm traceless}.  \label{eq:defineA}
\end{equation}
We shall use these variables as coordinates to parametrize the SU(N)
group, and we write $U_{x,i} = U(A_{x,i})$.  The manifold of
the SU(N) group requires more than one coordinate patch, but we shall not
trouble to account for this explicitly.  The formulas of the last
section hold, with
$\theta_{x,i}^\alpha = A_{x,i}^\alpha$.  
These coordinates agree with the exponential mapping, 
$U_{x,i} = \exp A_{x,i}$, to second order, for with the above definition
one may show
\begin{equation}
U_{x,i} = 1 + A_{x,i} + (1/2)(A_{x,i}^2) + (6N)^{-1} {\rm tr} A^3 +
O(A^4) ,
\end{equation}
so these variables approach the continuum connection $A_i(x)$ in the
continuum limit.

	In terms of these variables, the condition $dF_U(s) / ds|_{s=0} = 0$
reads
\begin{equation}
\sum_x {\rm tr} [ \omega_x \sum_i ( - A_{x,i} + A_{x-\hat{i},i}) ] = 0 .
\end{equation}
Since this holds for arbitrary $\omega$, we conclude that, at a minimum
of the minimizing function, $A$ satisfies
\begin{equation}
  \sum_i \; (A_{x,i} -  A_{x-\hat{i},i} ) = 0 .
\end{equation}
This is the lattice analog of the transversality condition that
characterizes the continuum Coulomb gauge, and we call our gauge choice
the ``minimal lattice Coulomb gauge''.  	The lattice Coulomb-gauge
condition is linear in the variables
$A_{x,i}$, and it may be solved by lattice fourier transform.

	To see the geometrical meaning of the gauge condition, and for future
use, it is helpful to introduce some definitions. 
We define the lattice ``gradient'' by
\begin{equation}
(\nabla \omega)_{x,i} \equiv (\omega_{x+\hat{i}} - \omega_x),
\end{equation}
It is a matrix that linearly maps site variables into link variables.
The identity, 
\begin{equation}
\sum_{x,i} {\rm tr} [ ( \omega_{x+i} - \omega_x ) A_{x,i} ]
= \sum_x {\rm tr} [ \omega_x \sum_i ( - A_{x,i} + A_{x-\hat{i},i}) ],
\end{equation}
holds for all $\omega$ and $A$.  We write it
\begin{equation}
(\nabla \omega, A) = (\omega, \nabla^* A)
\end{equation}
which defines the dual $\nabla^*$ of the lattice gradient.  
It is a matrix that linearly maps link variables into site variables,
that is given explicitly by
\begin{equation}
(\nabla^* A)_x = - \sum_i (A_{x,i} - A_{x-\hat{i},i)} ) ,
\end{equation}
and we have $\nabla^* = $ - (lattice ``divergence'').
These geometric
definitions may be extended to arbitrary lattices that need not be cubic
or periodic \cite{Z2}. In the minimal Coulomb gauge, the lattice
divergence of A vanishes, $ (\nabla^* A)_x = 0$.

	At a minimum of the minimizing function we also have the condition 
$d^2 F_U(s) / ds^2|_{s=0} \geq 0$.  From (\ref{eq:dFds}) we obtain  
\begin{equation}
d^2F_U(s) / ds^2 = (2N)^{-1} \sum_{x,i} \;  {\rm tr} 
\{ [ \omega_{x+\hat{i}} - \omega_x ] 
(d/ds)[( \, ^gU)_{x,i} - ( \, ^gU)_{x,i}^\dagger \; ] \: \} .
\end{equation}
With
\begin{equation}
A_{x,i}(s) \equiv A[ (^{g(s)}U)_{x,i} ] ,
\end{equation}
the positivity condition reads
\begin{equation}
d^2F_U(s) / ds^2|_{s=0} = N^{-1} \sum_{x,i} \;  {\rm tr} 
\{ [ \omega_{x+\hat{i}} - \omega_x ] \;
dA_{x,i}(s) / ds \: \}|_{s=0} \geq 0.
\end{equation}

	To make explicit the geometrical meaning this condition, and for
future use, we define the lattice gauge covariant ``derivative''
$D(A)$ by
\begin{equation}
[D(A)\omega]_{x,i} \equiv - dA_{x,i}(s) / ds|_{s=0}. 
\label{eq:defineD}
\end{equation}
It is the matrix that maps an infinitesimal gauge transformation
$\omega$ into the corresponding first-order change in the coordinates. 
(The minus sign is to be coherent with the continuum formula 
$^gA = A - D(A)\omega$.)  Like the lattice gradient $\nabla$, it maps
site variables into link variables, and we have $D(1) = \nabla$.  Its
dual $D(A)^*$ maps link variables into site variables, and is defined by
$(D^*(A)V,\omega) \equiv (V, D(A)\omega)$, for arbitary link variables $V$
and site variables $\omega$.  For the coordinates
$A$ defined above,
$D(A)$ is given explicitly by
\begin{equation}
[D(A)\omega]_{x,i} = t^b [D(A)\omega]_{x,i}^b
= (1/2) [ (\: U_{x,i} \omega_{x+\hat{i}} - \omega_x U_{x,i} )
  - ({\rm h.c.})]_{\rm traceless},
\end{equation} 
where $U = U(A)$.

	The positivity condition $d^2F_U(s)/ds^2|_{s = 0} \geq 0$, which holds at
absolute or relative minima of the minimizing function, thus reads
\begin{equation}
\sum_{x,i} (\nabla \omega)_{x,i}^b (D(U) \omega)_{x,i}^b \geq 0
\; {\rm for \: all} \; \omega .
\end{equation}
We define the lattice Faddeev-Popov matrix
\begin{equation}
M(A) \equiv D^*(A)\nabla,
\end{equation}
which maps site variables into site variables.  For transverse $A$, one
may verify that this matrix is symmetric,
\begin{equation}
D(A)^*\nabla = \nabla^* D(A) \;\;\; {\rm for} \; \nabla^*A = 0.
\end{equation}
We conclude that at a minimum of the minimizing function, this matrix is
also non-negative
\begin{equation}
M(A) \geq 0 \;\;\; {\rm for} A \in \Lambda.
\end{equation}  
It has a trivial null space consisting of $x$-independent eigenvectors,
$\nabla \omega = 0$.  This is a reflection of the fact that the minimal
Coulomb gauge does not fix global ($x$-independent) gauge
transformations.  On the orthogonal subspace, $M(A)$ is strictly positive
for configurations $A$ that are interior points of the fundamental
modular region $\Lambda$.  (For additional properties of
$\Lambda$, see \cite{Z2}.)

\section{Lattice Coulomb-gauge Hamiltonian}

In the preceeding section we obtained a parametrization of the
gauge orbit space by means of transverse configurations restricted to the
fundamental modular region.  We shall now calculate the lattice
Coulomb-gauge Hamiltonian $H_{\rm coul}$ as the restriction of the
Kogut-Susskind Hamiltonian $H_{\rm KS}$ to gauge-invariant wave
functions, $\Phi( ^gU) = \Phi(U)$.

	We represent states $\Phi(U)$ as functions of the coordinates, 
$\Phi = \Phi(A)$.  The coordinates transform according to $^gA_{x,i}^b =
A_{x,i}^b( ^gU)$ under local gauge transformation $g_x$.  For an
infinitesimal gauge transformation
$g_x = \exp(\omega_x) = 1 + \omega_x$, we have  
\begin{equation}
^gA_{x,i}^b = A_{x,i}^b - [D(A)\omega]_{x,i}^b ,
\end{equation}
by the definition
(\ref{eq:defineD}) of the lattice gauge-covariant derivative $D(A)$. 
With $\Phi( ^gA) = \Phi(A - D(A)\omega)$,  we conclude that
gauge-invariant wave functions, which satisfy
$\Phi( ^gA) = \Phi(A)$ for all $g$, satisfy the first order differential
equation,
\begin{equation} 
 \sum_{x,i} (D(A)\omega)_{x,i}^a \partial \Phi / \partial A_{x,i}^a = 0,
\end{equation}
for all $\omega$.  This is equivalent to the condition
\begin{equation}
(D^*(A)  \partial / \partial A)_x^b  \Phi = 0
\end{equation}
for all lattice sites $x$, which is the lattice version of Gauss's law
constraint.  This equation holds for any coordinates $A_{x,i}$
on the group manifold, provided only that the lattice gauge-covariant
derivative $D$ is  defined by (\ref{eq:defineD}).   

	We now proceed as in the continuum theory, and introduce the lattice
Faddeev-Popov identity
\begin{equation}
1 = \int \prod_x dg_x {\prod_x}' \delta_\Lambda[({\nabla^*} \, ^gA)_x] 
\det M'( ^gA).
\end{equation} 
The $\delta$-function
$\delta_\Lambda$ is defined by strict analogy to the continuum theory. 
The product $\prod_x dg_x$ extends over all sites $x$ of the lattice,
where $dg_x$ represents Haar measure.  However the primed product,
${\prod_x}' \delta_\Lambda[(\nabla^*A)_x]$ extends over all but one site
$x_0$ of the lattice because the lattice divergences satisfy the identity
$\sum_x (\nabla^*A)_x = 0$, and are thus not all linearly independent. 
The primed matrix $M'(A)$ is the lattice Faddeev-Popov matrix
$M(A) = D^*(A)\nabla$, with the rows and columns labelled by $x_0$
deleted.  One may show that the integrand is independent of $x_0$ and that,
apart from an overall normalization constant, $\det M'(A)$ for $\nabla^*A
= 0$ is the determinant of $M(A)$ on the space orthogonal to its
trivial null space.  In the minimal Coulomb gauge
$M'(A)$ is a strictly positive matrix, in the interior of the
fundamental modular region $\Lambda$, and its determinant is positive. 

	[The above Faddeev-Popov identity also holds for other coordinates,
such as that provided by the exponential mapping 
$U_{x,i} = \exp A_{x,i}$, with gauge condition $(\nabla^*A)_x = 0$.
We now
derive an explicit expression for the lattice gauge-covariant derivative 
$D(\theta)$ in any
coordinate system
$\theta_{x,i}^\alpha$, where $D(\theta)$ is defined by the linear change
in the coordinates induced by an infinitesimal local gauge
transformation.  We start with the Maurer-Cartan differential
\begin{equation}
U_{x,i} dU_{x,i}^\dagger = 
t^a \psi_\alpha^a(\theta_{x,i}) d\theta_{x,i}^\alpha,
\end{equation} which relates infinitesimal changes in the coordinates
$d\theta$ to infinitesimal changes $dU$ in $U$.  For an infinitesimal
gauge transformation
$g_x = 1 + \omega_x$ we have
\begin{eqnarray}
^gU_{x,i} & = & U_{x,i} + \omega_x U_{x,i} - U_{x,i} \omega_{x+\hat{i}}
\nonumber \\
dU_{x,i} & = & \omega_x U_{x,i} - U_{x,i} \omega_{x+\hat{i}}  \nonumber \\
dU_{x,i}^\dagger & = & - U_{x,i}^\dagger \omega_x 
+ \omega_{x+\hat{i}}U_{x,i}^\dagger  \nonumber \\
U_{x,i}dU_{x,i}^\dagger 
& = & U_{x,i}\omega_{x+\hat{i}}U_{x,i}^\dagger - \omega_x,
\end{eqnarray}
where we have used $\omega_x^\dagger = - \omega_x$.
The last expression represents the difference between the
parallel-transport of $\omega$, from $x+\hat{i}$ to $x$, and $\omega_x$. 
By analogy with the continuum theory, it is natural to call this quantity
the lattice gauge-covariant derivative in the {\em site}
 basis,
\begin{equation}
[{\cal D}(U)\omega]_{x,i} \equiv 
U_{x,i}\omega_{x+\hat{i}} U_{x,i}^\dagger - \omega_x .
\end{equation}
If we choose a basis in the Lie algebra, $\omega_x = t^a \omega_x^a$, we
have 
$U_{x,i} t^b U_{x,i}^\dagger = t^a R^{ab}(U)$,
where $R^{ab}(U)$ is the adjoint representative of $U$.  In the same Lie
algebra basis we have the expansion
$[{\cal D}(U)\omega]_{x,i} = t^a [{\cal D}(U)\omega]_{x,i}^a$.
This gives the explicit expression for the lattice gauge-covariant
derivative in the site basis
\begin{equation}
[{\cal D}(U)\omega]_{x,i}^a = 
R^{ab}(U) \omega_{x+\hat{i}}^b - \omega_x^a  .
\end{equation}
By analogy with the continuum formula for an infinitesimal gauge
transformation $^gA = A - D(A)\omega$, we define the lattice
gauge-covariant derivative in the {\em link} basis
$D\omega$ as the first order change in the link coordinates induced by
an infinitesimal gauge transformation 
\begin{equation}
d\theta_{x,i}^\alpha \equiv [D(\theta)\omega]_{x,i}^\alpha \; .
\end{equation}
The two quantities are related by use of the Maurer-Cartan differential
\begin{equation}
t^a \psi_\alpha^a(\theta_{x,i}) d\theta_{x,i}^\alpha =
U_{x,i}dU_{x,i}^\dagger = t^a [{\cal D}(U)\omega]_{x,i}^a \;
\end{equation}
which gives
\begin{equation}
\psi_\alpha^a(\theta_{x,i}) [D(\theta)\omega]_{x,i}^\alpha =
[{\cal D}(U)\omega]_{x,i}^a \; .
\end{equation}
The last formula may be inverted using
$J_a^\beta (\theta_{x,i}) \psi_\alpha^a(\theta_{x,i}) 
= \delta_\alpha^\beta$, 
\begin{equation}
[D(\theta)\omega]_{x,i}^\alpha = 
J_a^\alpha(\theta_{xi}) [{\cal D}(\theta)\omega]_{x,i}^a \; ,
\end{equation}
and we obtain for the lattice gauge-covariant
derivative in the link basis the explicit formula, valid in any
coordinate system
\begin{equation}
[D(\theta)\omega]_{x,i}^\alpha = 
J_a^\alpha(\theta_{xi}) 
[R^{ab}(\theta) \omega_{x+\hat{i}}^b - \omega_x^a ] \; ,
\end{equation}
where we have written $R^{ab}(\theta) = R^{ab}(U(\theta))$.]

	We shall now derive the lattice Coulomb-gauge Hamiltonian $H_{\rm coul}$
in any coordinate system with gauge condition 
$\sum_{x,i} (A_{x,i} - A_{x-\hat{i}})$.  We insert the Faddeev-Popov
identity into the formula for the inner-product $(\Phi, \Psi)$.  Because
the Haar measure 
$dU_{x,i} = \det\psi(A_{x,i}) dA_{x,i}$ is invariant
under left and right group multiplication, it is invariant under local
gauge transformation.  Consequently we obtain, just as in the continuum
theory,
\begin{equation}
(\Phi, \Psi) = N \int_\Lambda \prod_{x,i} dA_{x,i} \psi(A_{x,i})
{\prod_x}'\delta[(\nabla^* A)_x] \det M'(A) \Phi^*(A) \Psi(A),
\end{equation}
where N is the finite volume of the local gauge group, and we have
written $\psi$ for $\det \psi$.  The primed product ${\prod_x}'$
extends over all sites but one, as explained in the preceding section,
and correspondingly for $M'$. 

 	The gauge condition $\nabla^*A = 0$ may be
solved by a fourier decomposition of $A_{x,i}$ on a finite, periodic,
cubic lattice, using longitudinal and transverse polarization vectors
for finite $k$, and including $k = 0$ or harmonic modes.  This allow
to write
\begin{equation}
A_{x,i,\alpha} = A_{x,i,\alpha}^{\rm Tr} 
- (\nabla \sigma_\alpha)_{x,i},
\end{equation}
where $(\nabla^*A_\alpha^{\rm Tr})_x = 0$.  Here $A^{\rm Tr}$ includes
both transverse and harmonic modes, 
$A^{\rm Tr} = A^{\rm tr} + A^{\rm h}$. The integral over link variables
which represents the inner product becomes an integral over fourier
coefficients. Because of $\delta(\nabla^*A)$, only the transverse and
harmonic modes survive.  We express the result as 
\begin{equation}
(\Phi,\Psi) = N \int_\Lambda dA^{\rm Tr} \prod_{x,i} 
\psi(A_{x,i}^{\rm Tr}) \det M'(A^{\rm Tr}) 
\Phi^*(A^{\rm Tr}) \Psi(A^{\rm Tr}),
\end{equation}
where $dA^{\rm Tr}$ designates the interal over all transverse and
harmonic modes.

	We wish to express the Hamiltonian $H_{KS} = T + V$ as an operator on the
reduced space of gauge orbits parametrized by $A^{\rm Tr} \in \Lambda$. 
The potential energy operator
$V$ acts in the larger space by multiplication by a gauge-invariant
function $V( ^gA) = V(A)$.  On the reduced space it acts simply by
multiplication by $V = V(A^{\rm Tr})$.  

The Lie generator
$J_a$ is symmetric with respect to Haar measure, so we may write the
kinetic energy operator
$T$ in the Kogut-Susskind representation as the quadratic form
\begin{equation}
(\Phi, T\Psi) = \int \prod_{x,i} dU_{x,i} (g_0^2/2)
\sum_{x,i, a} ({\cal E}_{x,i,a} \Phi)^* \;  ({\cal E}_{x,i,a} \Psi).
 \end{equation} 	
Only first derivatives of the wave-function appear here, which will allow
us to apply Gauss's law as in the continuum theory.   Here we have written
\begin{equation}
{\cal E}_{x,i,a} \equiv J_{x,i,a} 
\end{equation}
in order to emphasize the analogy with the continuum electric
field.  The operator ${\cal E}_{x,i,a}$ represents the electric field in
the the site basis.  We also introduce the electric field in the link
basis.  
\begin{equation}
E_{x,i,\alpha} \equiv i \partial / \partial A_{x,i}^\alpha .
\end{equation}
The two are related by
\begin{equation}
{\cal E}_{x,i,a} = J_a^\alpha(A_{x,i}) E_{x,i,\alpha} .
\end{equation}
In this notation, Gauss's law constraint reads
\begin{equation}
(D(A)^* E)_x \Phi = 0.
\end{equation}

	The operator ${\cal E}_{x,i,a} = J_{x,i,a}$ transforms homogeneously
under gauge transformation, so the integrand of the quadratic
form $T$ is gauge invariant.  We insert the lattice Faddeev-Popov
identity into this integral, and obtain
\begin{eqnarray}
(\Phi, T\Psi) & = & N \int_\Lambda  \prod_{x,i} dA_{x,i} \: 
\psi(A_{x,i}) \det M'(A)  \nonumber \\
&  & \times {\prod_x}' \delta[(\nabla^* A)_x]
(g_0^2/2) \sum_{x,i, a} ({\cal E}_{x,i,a} \Phi)^* \;  
({\cal E}_{x,i,a} \Psi).
\end{eqnarray}

	We wish to use the Gauss's law constraint to express $\cal E$ as a
derivative operator that acts only on transverse variables.  For this
purpose we make a Fourier decomposition of the electric field
$E_{x,i,\alpha}$ on the periodic cubic lattice. For the
$A_{x,i}^\alpha$-field we write the fourier decomposition,
as
\begin{equation}
A_X = \sum_K \phi_{X,K} \tilde{A}_K,
\end{equation}
where $X$ represents position and vector indices, and $K$ represents
wave vector and polarization indices.  Here the $\phi_{X,K}$ form a real
orthonormal basis, and the inversion reads
\begin{equation}
\tilde{A}_K = \sum_X \phi_{X,K} A_X.
\end{equation} 
We have
\begin{equation}
\frac{\partial \; \; \; \; } {\partial A_X} = 
\sum_K \frac{\partial \tilde{A}_K}{\partial A_X} 
\; \frac{\partial \; \; \; \; }{ \partial \tilde{A}_K} 
=\sum_K \phi_{X,K} \; \frac{\partial \; \; \; \;} {\partial \tilde{A}_K},
\end{equation}
so the fourier decompositions of $A_X$ and of
$\frac{\partial \; \; \; \; } {\partial A_X}$ are the same.  For the
electric field $E_X$ we make the decomposition onto the same basis
\begin{equation}
E_X = \sum_K \phi_{X,K} \tilde{E}_K .
\end{equation} 
Consequently, from 
\begin{equation} 
E_X = i\frac{\partial \; \; \; \; } {\partial A_X}, 
\end{equation}
we obtain
\begin{equation} 
\tilde{E}_K = i\frac{\partial \; \; \; \; } {\partial \tilde{A}_K}.
\end{equation}
We choose a polarization basis which is adapted to the decomposition into
transverse and longitudinal and harmonic parts.  Corresponding to the
decomposition 
$A = A^{\rm Tr} - \nabla \sigma$, we have
\begin{equation}
E = E^{\rm Tr} - \nabla \Omega,
\end{equation}
where $\nabla^*A^{\rm Tr} = \nabla^*E^{\rm Tr} =0$, and we have
introduced the lattice color-Coulomb potential operator $\Omega$.  It
follows that if
$A^{\rm Tr}$ has the fourier decomposition
\begin{equation}
A_X^{\rm Tr} = \sum_{K \in T} \phi_{X,K} \tilde{A}_K,
\end{equation} 
then $E^{\rm Tr}$ has the fourier decomposition
\begin{equation}
E_X^{\rm Tr} = i \sum_{K \in T} \phi_{X,K} \frac{\partial \; \; \; \; }
{\partial \tilde{A}_K},  \label{eq:expandET}
\end{equation}
where $T$ is the set of transverse and harmonic modes $K$.  This
expression defines $E^{\rm Tr}$ as a differential operator that acts
within the space of functions of $A^{\rm Tr}$.

	We shall now use the Gauss's law constraint $(D^*E)_x \Phi = 0$ to
determine the color-Coulomb potential $\Omega_x \Phi$ on the reduced
subspace, parametrized by $A^{\rm Tr} \in \Lambda$.  We write Gauss's law
in the form
\begin{equation}
D^*(E^{\rm Tr} - \nabla \Omega) \Phi = 0,
\end{equation}
which gives
\begin{equation}
D^* \nabla \: \Omega \: \Phi = \rho \: \Phi,
\end{equation}
where
\begin{equation}
(\rho)_x \equiv (D^*E^{\rm Tr})_x = [(D - \nabla)^*E^{\rm Tr}]_x \; .
\end{equation} 
This operator represents the color-charge density carried by the dynamical
degrees of freedom of the gluons.   
We shall only need to solve Gauss's law for $\Omega \Phi$ in the reduced
space where the Faddeev-Popov matrix is symmetric in the minimal Coulomb
gauge,
$M(A^{\rm Tr}) = D^*(A^{\rm Tr}) \nabla = \nabla^*D(A^{\rm Tr})$.
Thus we have also 
\begin{equation}
(\nabla^*D  \: \Omega)_x \: \Phi = \rho_x \: \Phi .
\end{equation}
Because
$\sum_x (\nabla^* V)_x = 0$ for any link variable $V$, the last equation
is consistent only if the total charge 
$Q \equiv \sum_x \rho_x$ of the state $\Phi$ vanishes. Thus for
gauge-invariant states in the minimal Coulomb gauge we have the
constraint on the reduced space
\begin{equation}
Q \Phi = 0.
\end{equation}

	In the preceding section we have seen that in the lattice Coulomb gauge
$D^*(A^{\rm Tr}) \nabla$ is also a strictly positive matrix for 
$A^{\rm Tr}$ in the interior of the fundamental modular region
$\Lambda$. Consequently the equation for $\Omega$ always has a solution,
\begin{equation}
\Omega \: \Phi = (D^* \nabla)^{-1} \rho \: \Phi \; .
\end{equation}
With $\rho = D^*(A^{\rm Tr})E^{\rm Tr}$, this expresses the color-Coulomb
potential operator $\Omega$ on the reduced space as a derivative with
respect to the components of
$A^{\rm Tr}$.  We conclude that in the reduced space the color-electric
field $E = E^{\rm Tr} - \nabla \Omega$ acts according to
\begin{equation}
E \Phi = E^{\rm Tr} \; \Phi - \nabla (D^* \nabla)^{-1} D^* E^{\rm Tr}
\; \Phi,
\end{equation}
where
$E^{\rm Tr}$ is defined by its fourier expansion (\ref{eq:expandET}). 

	With this result we may express the kinetic
energy operator $T$ as a quadratic form on the reduced space,
\begin{eqnarray}
(\Phi, T\Psi) & = & N \int_\Lambda  dA^{\rm Tr} 
\prod_{x,i}  \: \psi(A_{x,i}^{\rm Tr}) \det M'(A^{\rm Tr})  \nonumber \\
&  & \times (g_0^2/2) \sum_{x,i, a} ({\cal E}_{x,i,a} \Phi)^* \;  
({\cal E}_{x,i,a} \Psi),
\end{eqnarray}
where $\Phi = \Phi(A^{\rm Tr})$ and similarly for $\Psi$.  Here 
${\cal E} \Phi$ is defined by 
${\cal E}_{x,i,a} \Phi \equiv J_a^\alpha(A_{x,i}^{\rm Tr})
E_{x,i,\alpha}\Phi$, and $E\Phi$ is defined by the preceding equation. 
This completes the specification of the lattice Coulomb-gauge
Hamiltonian, $H_{\rm coul} = T + V$.

\section{Conclusion}

	We have derived by elementary methods the lattice Coulomb-gauge
Hamilonian $H_{\rm coul}$ from the Kogut-Susskind Hamiltonian 
$H_{\rm KS}$.  The essential step was to write $H_{\rm KS}$ as a quadratic
form, involving only first derivatives, so Gauss's law
$D^*E \Phi = 0$ could be use directly to solve for the color-Coulomb
potential.

	We have not addressed here invariance with respect to finite gauge
transformations, which involves identification of points
of the boundary of the fundamental region $\Lambda$, so that Gribov
copies are avoided.  These questions are addressed in \cite{Z2}, and
have an important influence on the spectrum \cite{Luescher,
Cutkosky, vanBaal, vdHeuvel}.

	However we believe that Gauss's law is essential to the phenomenon
of confinement in QCD \cite{Z1}.  We have shown how, in
Hamiltonian theory with lattice regularization, the imposition gauge
invariance in the form of Gauss's law leads to an instantaneous
color-Coulomb potential in the lattice Coulomb-gauge Hamiltonian $H_{\rm
coul}$.

\section{Acknowledgments}  I wish to express my appreciation to the NATO
Advanced Study Institute for its support, and to Pierre van Baal whose
leadership and tireless efforts made the workshop a success.


\end{document}